# SEMI SYMMETRIC METHOD OF SAN STORAGE VIRTUALIZATION

Mrs. Dhanamma Jagli[1], Mr. Ramesh Solanki1[2], Mrs. Rohini Temkar[3],

Mrs. Laxmi Veshapogu [4]

[1,2, 3]Assistant Professor, Department of MCA,
V.E.S. Institute Of Technology,Mumbai-74, India.
[1]dhana1210@yahoo.com,
[2]ram_solanki@rediffmail.com,
[3]rohini_dighe@rediffmail.com,
[4]Assistant Professor, Department of ECE,
Bharati Institute of Technology,Hyderabad
[4]nakirekantilaxmi@yahoo.com

## Abstract

*Virtualization is one of the biggest buzzwords of the technology industry right at this moment. The fast growth in storage capacity and processing power in enterprise installations coupled with the need for high availability, requires Storage Area Network (SAN) architecture to provide seamless addition of storage and performance elements without downtime. The usual goal of virtualization is to centralize administrative tasks while improving scalability and work loads. This paper, describing about new proposed method for virtualization, which would be overcome limitations of existed methods for storage virtualization.*

## I. Introduction

These days, there is significant interest in developing more agile data centers, in which applications are loosely coupled to the underlying infrastructure and can easily share resources among themselves. Also desired is the ability to migrate an application from one set of resources to another in a nondestructive manner. Such agility becomes key in modern cloud computing infrastructures that aim to efficiently share and manage extremely large data centers. One technology that is set to play an important role in this transformation is virtualization [5].

Current data centers require storage capacities of hundreds of terabytes to petabytes. Time-critical applications such as on-line transaction processing depend on getting adequate performance from the storage subsystem; otherwise, they fail. It is difficult to provide predictable quality of service at this level of complexity, because I/O workloads are extremely variable and device behavior is poorly understood. Ensuring that unrelated but competing workloads do not affect each other's performance is still more difficult and equally necessary [1].

Virtualization is the pooling of physical storage from multiple network storage devices into what appears to be a single storage device that is managed from a central console Virtualization is the act of integrating one or more (back-end) services or functions with additional (front-end) functionality for the purpose of providing useful abstractions [2].Storage virtualization is commonly used in a storage area network (SAN). The management of storage devices can be tedious and time-consuming. Storage virtualization helps the storage administrator perform the tasks of backup, archiving, and recovery more easily, and in less time, by disguising the actual complexity of the SAN [9].

## II. Related work

### a) Virtualization Technologies

Virtual Machine technology, first introduced in the 1960s, has been widely exploited in recent years for consolidating hardware infrastructure in enterprise data centers with technologies like VMware and Xen .While server virtualization has garnered a lot of attention from the academic and research communities, storage virtualization has received considerably less attention. Storage virtualization refers to the process of abstracting physical storage into virtualized containers called *virtual disks* (*Vdisks*) that can be used by applications.

### b) Storage Area Network (SAN)

A storage area network (SAN) is architecture to connect detached computer storage devices, such as disk arrays, tape libraries, and optical jukeboxes, to servers in a way that the devices appear as local resources. SANs deliver storage to servers at a block level, and feature mapping and security capabilities to ensure only one server can access the allocated storage at any particular time. The protocol, or language, used to communicate between storage devices and servers is SCSI (small computer system interface).Traditionally, SANs have used optical connectivity called fiber channel (FC) due to its gigabit transfer speeds and ability to work over long distances. However, more recently Ethernet based storage networks have become popular due to the ubiquity (and therefore lower cost) of the technology in the TCP/IP world (transmission control protocol/Internet protocol) and the adoption of the SCSI protocol, called iSCSI (Internet small computer system interface) to the medium [1].

The storage area network in the data center is composed of servers (hosts), switches and storage subsystems connected in a hierarchical fashion mostly through a FibreChannel (FC) [8] network fabric. Each server has one or more Host Bus Adapters (HBAs) with one or more FibreChannel ports each. These ports are connected to multiple layers of SAN switches which are then connected to ports on storage subsystems. In a large enterprise data center, there can be as many as thousands of servers, hundreds of switches and hundreds of storage subsystems [5].

### c) Network Attached Storage (NAS)

Network attached storage (NAS) has traditionally been a single device that contains disk storage and computer components (CPU, memory, network ports) and whose sole purpose is to deliver storage to servers. NAS is sometimes referred to as a 'file server' due to its ability to share a common pool of files with multiple servers. By managing the disks, file systems, and volumes, it optimizes and simplifies the delivery of storage to servers in a network environment, and can provide services such as redundancy and replication, relieving servers of these tasks.NAS storage typically connects to an organization's TCP/IP network via Ethernet (the same used by servers) and uses specific languages, or protocols, like NFS (network file system) or CIFS (common Internet file system), based on the server message block protocol, to exchange information between servers. More recent implementations of NAS appliances use SANs to access attached block-based storage.

### d) Storage Virtualization

With the introduction of commercial computers back in the 1950s, like IBM's mainframes, the magnetic disk storage used to maintain information was directly attached via a cable, or a *bus,* to the computer's central processing unit (CPU) and volatile random access memory (RAM). This method of attachment, commonly referred to as *direct attached storage (DAS)* , provided a modest amount of non-volatile storage to a single computing system. Over the course of the next few decades, standard *protocols* (the language used to communicate between devices) emerged, such as the small computer system interface (SCSI), making it easy. Storage virtualization can be broken up into two general classes: block virtualization and file virtualization. Block virtualization is best summed up by Storage Area Network (SAN) and Network Attached Storage (NAS) technologies: distributed storage networks that appear to be single physical devices. Under the hood, SAN devices themselves typically implement another form of Storage Virtualization: RAID. iSCSI is another very common and specific virtual implementation of block virtualization, allowing an operating system or application to map a virtual block device, such as a mounted drive, to a local network adapter (software or hardware) instead of a physical drive controller. The iSCSI network adapter translates block calls from the application to network packets the SAN understands and then back again, essentially providing a virtual hard drive. File virtualization moves the virtual layer up into the more human-consumable file and directory structure level. Most file virtualization technologies sit in front of storage networks and keep track of which files and directories reside on which storage devices, maintaining global mappings of file locations.

e) **Data virtualization**

It is about placing all your diverse physical database assets behind Technology that provides a single logical interface to your data. Data is more logically grouped to make that data more useful, which this has been the promise of SOA for years. An SOA with a data services layer is a valid architectural approach to dealing with both underlying business processes as well as sets of services. Over time, many other data sources get added or removed, and information residing within specific applications throughout the enterprise becomes so complex that it's difficult to figure out how key business entities such as customer, sales, and inventory reside within the physical data sources. The complexity makes the core data sources almost unusable.

Data virtualization reduces the complexity by creating sets of data services or a data abstraction layer that maps from the complex physical data source structures and data, to core business entities, into a virtual database. Data virtualization can help by enabling better data governance. Data virtualization is a real-time data integration approach. Data is accessed on demand, directly from the data sources, eliminating additional processing or staging of the data.

**Storage Virtualization Benefits [1]**
1. Optimizing Performance and Improving Storage Utilization
2. Understanding thin, or dynamic, provisioning,
3. Creating a dynamic provisioning pool
   - ✓ Closely coupling storage purchases to the application's consumption.
   - ✓ Repurposing unused storage that is already allocated, using the storage only when the host (as per the application's demand) writes the data on it.
   - ✓ Simplifying storage management tasks by managing capacity from a central pool instead of server by server.
   - ✓ Reducing capital costs, as over-provisioning is reduced or eliminated.
   - ✓ Improving performance, as data sets are striped over a large number of drives.
4. Space saving
5. Wide striping for performance improvements.

# III. Available Methods for Storage Virtualization

## A. Virtualization at the Server Level

One method of virtualization is via storage management software that runs at the server level. The main advantage of this method is that it enables multiple storage subsystems to work in parallel with multiple servers. A key difficulty with this method is that it assumes a prior partitioning of the entire SAN resources to the various servers. Virtualization is only performed on pre-assigned storage, losing a key advantage of SANs as well as the independence of Volumes from Servers.

## B. Virtualization at the Storage Sub-system Level

This method was first implemented in mainframe environments in the 90's and is one of the most common storage virtualization solutions in use today. In this method, a uniform Storage Virtualization Manager is achieved by creating Virtual Volumes over the storage space of the specific storage subsystem. Pooling all SAN storage resources and managing Virtual Volumes across several storage subsystems requires that this method be augmented by other means which in general will only be practical in homogeneous SANs using a single type of RAID subsystem. Creating virtual volumes at the storage system level provides host independence, but with limited flexibility for growth.

### C. Symmetric Virtualization

Virtualization in a separate hardware box that is placed between the servers and the storage solves many of the difficulties of the above two approaches but at a very high price. The advantage of this method is that the Volumes are completely independent of the both the servers and the storage subsystems on the SAN. The management software "sees" all the physical storage available and can create virtual volumes and allocate them as required. The key drawback of this architecture is that it achieves this independence at a very high price i.e. every I/O of every server is sent through this central unit causing significant performance degradation and a SAN bottleneck. This method may thus lead to poor performance, expensive solutions, or both.

### D. Asymmetric Virtualization

This method uses a combination of a Metadata Center and Volume Drivers for creating and managing virtual volumes while enabling direct data transfer between server and storage subsystems. This method ensures negligible impact on SAN throughput, even for very large SANs. Indeed, by allowing multiple storage subsystems to work in parallel with multiple servers, total SAN performance can be considerably enhanced. But leads to redundancy of the Metadata Center and increasing maintenance cost. This method is more complex and every time needs to be search data path and extract information related to storage devices. Overall management of this method leads to overhead, new method is proposed.

## IV. Proposed System Architecture

### 1) Virtualization through Metadata centre

This method uses a grouping of a Metadata Center and Volume Drivers for managing virtual volumes. The virtualization software or device is responsible for maintaining a consistent view of all the mapping information for the virtualized storage. This information is usually called meta-data and is stored as a mapping table. Virtualization in a separate hardware box that is placed between the servers and the storage. This method consists of a Metadata Center that stores data about data at the common place. Volume Driver accesses the path data path from the Meta data centre and presents virtual volumes path to the every server by sending instruction to server. Server would access virtual volume from virtual hardware box which holds virtual volumes of storage sub systems as shown in Fig 1.

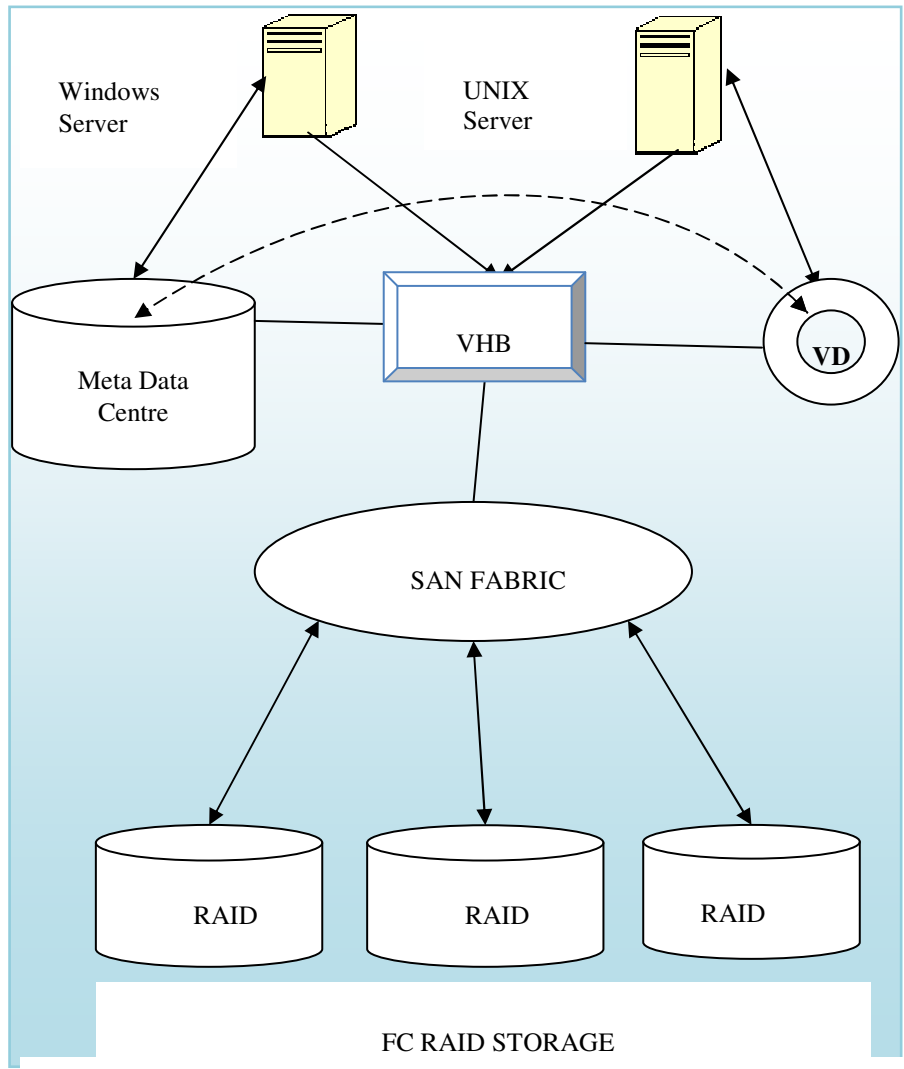

**Figure 1: Semi symmetric method**

In this method Metadata centre is used to provide access path to Volume Driver. Volume Driver is used to manage virtual configuration paths and maintain virtual data in the separate hard ware box. In this method perfqmance is increased, lowers the maintenance overhead and diminish the cost for management. This method providing advantages like flexibility, high scalability and single management is possible.

# V. Conclusion

Data virtualization solves the most intractable troubles facing in IT organizations. Data virtualization places an agile and configurable layer between back-end physical databases and the way these databases are represented using data services. This proposed method would be providing all the feature of existed storage virtualization methods.